# Moiré-induced lattice reconstruction at buried atomic interfaces

**Nicholas Clark[1,2†], Frederick Allars[3†], Isaac Soltero[1,2†], Wendong Wang[1,2†], David Hopkinson[3], Hugo de Latour[1,2], James G. McHugh[1,2], William Talbott[2,4], Sam Sullivan-Allsop[2,4], Rongsheng Cai[4], Astrid Weston[1,2], Xiao Li[1,2], Gareth Tainton[2,4], Casey K. Cheung[1,2], Francisco Selles[1,2], Alex Summerfield[2], Andrey Kretinin[2,4], Christopher S. Allen[3,5*], Vladimir Fal'ko[1,2*], Sarah J. Haigh[2,4,6*] and Roman Gorbachev[1,2,6*]**

[1]Department of Physics, University of Manchester; Manchester M13 9PL, UK
[2]National Graphene Institute, University of Manchester; Manchester M13 9PL, UK
[3]Electron Physical Science Imaging Centre, Diamond Light Source; Didcot OX11 0DE, UK
[4]Department of Materials, University of Manchester; Manchester M13 9PL, UK
[5]Department of Materials, University of Oxford; Parks Road, Oxford OX31 3PH, UK
[6]Henry Royce Institute for Advanced Materials; Manchester M13 9PL, UK

[†]These authors contributed equally to this work.

*Corresponding authors. Email: roman@manchester.ac.uk, sarah.haigh@manchester.ac.uk, vladimir.falko@manchester.ac.uk, christopher.allen@diamond.ac.uk.

**Atomic reconstruction at twisted two-dimensional interfaces governs many of their emergent optical, electronic, and mechanical properties, including sliding ferroelectricity. Despite recent progress in understanding lattice reconstruction in suspended twisted bilayers, structural changes at van der Waals heterointerfaces between multilayer crystals remain largely unexplored. Here we use multi-slice electron ptychography to non-invasively recover the three-dimensional atomic structure at marginally twisted rhombohedral interfaces between thick transition-metal dichalcogenide crystals. With a position precision of ~3 pm and a depth resolution ~1 nm, we resolve the twist-induced lattice reconstruction field per layer, and the resulting dislocation network at the buried interface. Despite the bulk nature, we observe markedly strong in-plane interfacial reconstruction due to suppression of the out-of-plane bending by outer layers, exceeding predictions from our three-dimensional modelling. Furthermore, we extract the strain tensor evolution during the decay of the reconstruction into the bulk, providing a structural foundation for understanding multi-layer moiré systems.**

Moiré superlattices form when crystalline surfaces are superimposed with a mismatch in rotational angle or lattice constant, and control of the twist angle and heterostructure composition alter their electronic, optical, mechanical, and thermal properties. Understanding the diverse phenomena arising in twisted interfaces requires detailed knowledge of their atomic structure. For instance, at small twist angles of a few degrees, the flexibility of 2D layers allows for lattice reconstruction,[1] leading to dramatic changes in their properties.[2] This phenomenon has been studied using surface-sensitive techniques such as scanning tunnelling microscopy,[3-6] atomic force microscopy (AFM),[1] scanning near-field optical microscopy,[7] piezoresponse force microscopy,[8] scanning electron microscopy (SEM)[9-11] and low-energy electron microscopy (LEEM) .[12] However, all these characterization methods require the reconstructed interface to be very near the surface. (Scanning) transmission electron microscopy [(S)TEM] can image domain structure through both direct visualization of the atomic structure[2] and reciprocal space imaging.[13] As a projection technique, it can be used on thicker samples such as encapsulated twisted bilayers[14,15] if the encapsulation layers are thin.

More complex and thicker systems, such as double bilayer graphene,[16-18] multilayer sliding ferroelectrics[19-21] and ferroelectric vortices in twisted multilayer $BaTiO_3$,[22] can exhibit lattice reconstruction at their buried mutual interface.[23,24] However, the extent of this phenomenon and the manner of its decay into the bulk is unexplored experimentally. Efforts have been made to infer the depth profile of the lattice reconstruction using scanning probe microscopy imaging of the outer surfaces in combination with modelling,[25-27] but true atomic imaging of such buried interfaces has not been achieved.

Studying such systems is particularly challenging because the atomic displacements are small. At the reconstructed interface, atoms move only by tens of picometres, and in the subsequent layers the displacements trail off to zero. When multiple layers are viewed in projection, tracking atomic displacements in the individual planes with conventional (S)TEM imaging modes becomes impossible because of the complexity of the overlaid atomic positions.

There are several (S)TEM techniques that attempt to resolve depth-dependent information. For example, use of a convergent electron beam produces a finite depth of focus allowing focal sectioning of three-dimensional (3D) structures,[22,28,29] but the depth resolution is limited to ~5 nm, which corresponds to >12 layers superimposed in a single slice. Recently, atomic electron tomography visualised the 3D structure of a monolayer $MoS_2$,[30] but its use for thicker and more complex twisted systems is yet to be explored.

Electron ptychography is a dose-efficient technique in which real-space phase information beyond the information limit set by diffractive optics[31] is extracted from analysis of overlapping scattered beams in diffraction patterns formed when a localised electron probe is scanned across the sample.[32-35] Most ptychographic approaches provide only the 2D projected phase map. However, multislice ptychography has emerged for recovery of depth-dependent atomic structure,[36-38] and has been successfully applied to oxides[39-43] with relatively large lattice parameters and non-reconstructed hBN films.[44]

We performed multislice electron ptychography to achieve 3D electron imaging of complex reconstructed interfaces between bulk transition metal dichalcogenide (TMD) crystals. We show that using a slice thickness corresponding to two TMD layers, we could individually track >40,000 atomic columns in the constituent bulk crystals. We extracted atomic displacements caused by interfacial reconstruction as a function of the distance from the interface with a precision as low as 3 pm comparing the results to reconstructed and rigid models. We supported our experimental results with customised theoretical modelling and generalised the outcomes for a wide range of twist angles and thicknesses. We observed substantial enhancement in the reconstruction of the multilayer moiré cell compared to the predicted behaviour due to the bending rigidity of the outer layers effectively "forcing" the inner layer to enhance in plane reconstruction.

**Initial image acquisition and ptychographic reconstruction**

We studied reconstruction at both homointerfaces between two thick $MoS_2$ crystals, and heterointerfaces between $MoS_2$ and $WS_2$ crystals, both with the favoured hexagonal (2H) internal stacking order. Samples were created by mechanically transferring the individual crystals on top of each other with a predefined twist angle of ~1-2°. The thickness of each piece was 12 to 18 atomic layers according to AFM measurements prior to the assembly. To minimise contamination, we developed a nanofabrication approach in which the specimen was transferred and supported using silicon nitride cantilevers[45] with lithographically defined windows for the transmission imaging (see Methods for fabrication and characterization details).

We acquired the ptychographic data sets by scanning a defocused electron probe across the sample. The transmitted diffraction patterns were recorded at each pixel position (Fig.1a). The convergence semi-angle was 31.7 mrad at an 80-kV accelerating voltage ensuring sufficient overlap between the diffracted and primary transmitted beam. An iterative multislice ptychographic reconstruction was performed using an ePIE algorithm[46,47] with a slice thickness corresponding to either 2 (1.23 nm) or 3 (1.84 nm) $MoS_2$ layers[48] depending on the dataset. In Fig. 1b, we show an example of the multislice ptychographic imaging of a twisted $MoS_2$-$MoS_2$ homointerface, with the middle slice, labelled "0", centred at the twisted interface, capturing the innermost layers of the two $MoS_2$ crystals. This 0th slice contained the strongest lattice reconstruction, as seen in Fig.1c, with extended domains of perfect MX' and XM' stacking forming in order to maximise the energetically favourable commensurate areas, similar to the results reported for a twisted $MoS_2$ bilayer.[2,49] In contrast, a slice positioned far from the twisted interface, Fig 1d, displayed a nearly perfect hexagonal lattice with equal intensities of alternating 2H stacked metal/chalcogen columns (slice $\underline{4}$ containing the 8th and 9th $MoS_2$ layers of the bottom crystal).

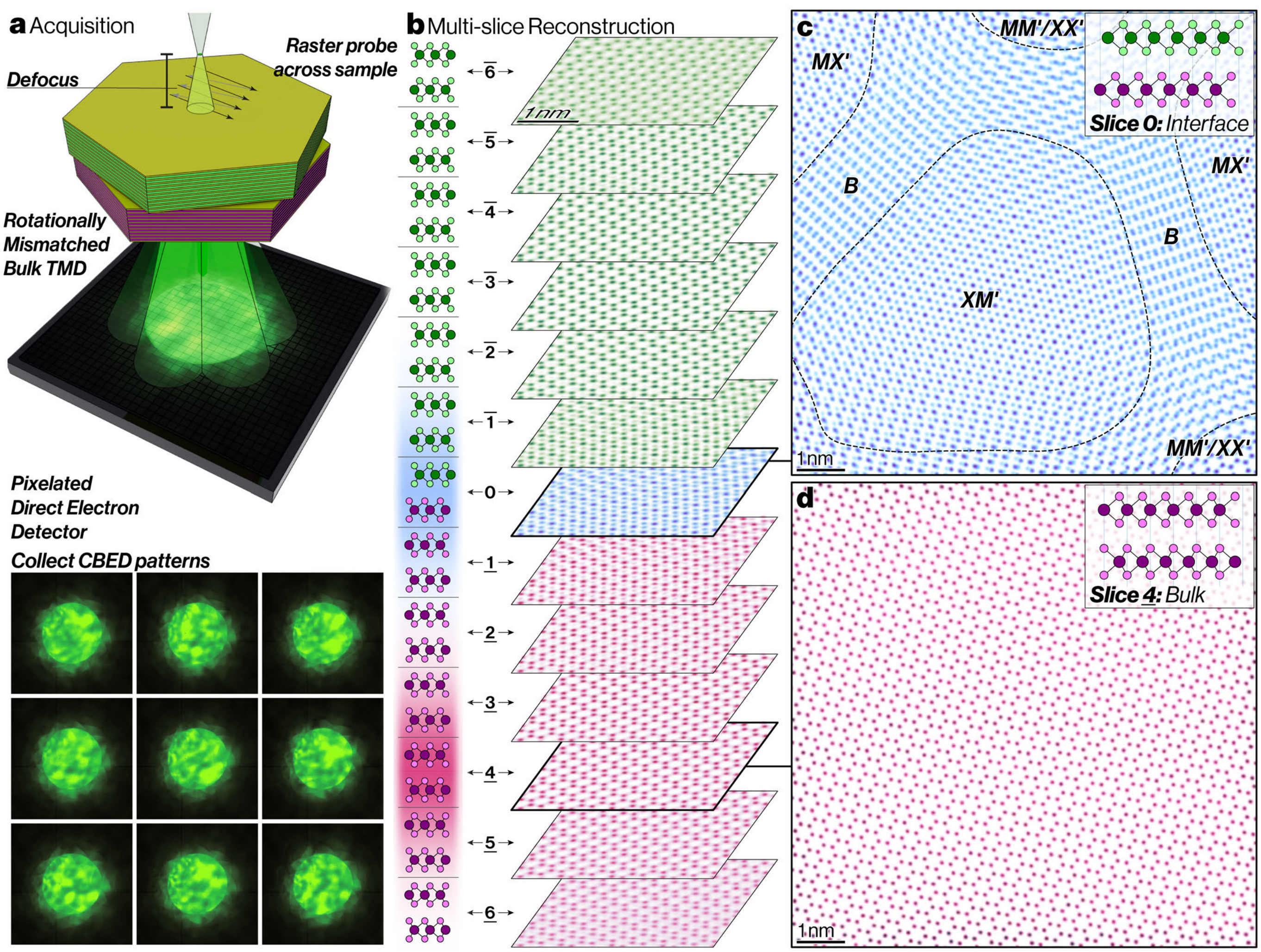

**Figure 1: Depth dependent atomic structure revealed through multislice ptychography (two $MoS_2$ layers per slice).** (**a**) Schematic showing 4D-STEM data collection for ptychographic reconstruction. The upper panel details the acquisition technique whereby a defocused converged probe is raster scanned over the sample surface with a real space pixel spacing smaller than the probe size at the sample. A convergent-beam electron diffraction (CBED) pattern is collected for each overlapping probe position; examples are shown in the lower panel. (**b**) Reconstructed phase slices extracted from the CBED map through multislice iterative ptychographic reconstruction, with each 1.22 nm thick slice corresponding to two $MoS_2$ layers. Slices are coloured red for the lower crystal, blue for the interface slice and green for the upper crystal. Left shows a schematic illustrating the presumed interlayer stacking within each phase slice, with slice 0 and $\underline{4}$ overlaid with a gaussian function representing the effective depth resolution. (**c**) and (**d**) show larger views of reconstructed phase slices within the lower crystal (d) and at the twisted interface (c). Reconstructed domains (MX', XM' and MM'/XX') are highlighted in (c), with the boundary drawn where the local interlayer displacement is within 50 pm of the relevant high symmetry stacking in each case.

### Subatomic displacements

To visualise the subatomic displacements, we analysed a different ptychographic dataset where the slices were positioned symmetrically on either side of a twisted homointerface, with each crystal containing 3 $MoS_2$ layers (1.8 nm per slice). This approach allowed us to directly compare how the atoms in the 1st slice on the "positive" side (top crystal denoted $\overline{1}$) are located with respect to the atoms in the 1st slice on the "negative" side (bottom crystal denoted $\underline{1}$). After tracking individual atoms in each slice, we calculated the displacement vector between the hollow centres in slice pairs (e.g. $\overline{2}/\underline{2}$, etc.). The resulting displacement maps are shown in Fig. 2b, with the displacement vector magnitude and direction colour-coded according to Fig. 2a.

The top panel ($\overline{1}/\underline{1}$) shows that those layers nearest to the twisted interface exhibited well-pronounced reconstruction and formed two types of extended domains (coloured in red and blue). In the 2H polytype (most energetically favourable for $MoS_2$ and $WS_2$), consecutive layers alternate in orientation by 180°, so the interface configuration is dependent on the relative orientation of its two innermost layers. Parallel aligned layers would generate an interface with nearly rhombohedral stacking (as found in 3R bulk crystals), whereas antiparallel alignment would generate a hexagonal twisted interface as in the 2H polytype. The two layer slice depth of the ptychographic data prevented us from discerning whether the twisted interface was nearly hexagonal (unit cells twisted by ≈180°) or nearly rhombohedral (unit cells twisted by ≈0°) by direct observation alone.

To distinguish between these two possible types of interfaces, we tracked the evolution of the black region, where the hexagonal lattices were perfectly aligned, with sample depth. If the twisted interface was nearly hexagonal, this region would represent the most favourable 2H stacking, which should dominate the interfacial area[2,49] at a twist angle of ~1°, contrary to our observations (direct comparison in Fig. S9). We concluded that the interface was nearly rhombohedral. The atomic pattern of the black regions in Fig. 2b represents the energetically unfavourable MM'/XX' stacking with the metal (M) Mo atoms in one layer aligned over the Mo in the other and the chalcogen (X) S atoms over S atoms. Consequently, the blue and red areas, labelled MX' and XM', where chalcogens in one layer were aligned over metal atoms in the other layer and vice versa, were the mirror-inverted pairs of rhombohedral domains that have broken inversion symmetry and which have been shown to host sliding ferroelectricity in bilayer form.[11] Further from the interface (slices $\overline{2}/\underline{2}$ and $\overline{3}/\underline{3}$) the blue and red areas gradually shrank and were similar in size to the black areas, as expected for a non-reconstructed moiré lattice.

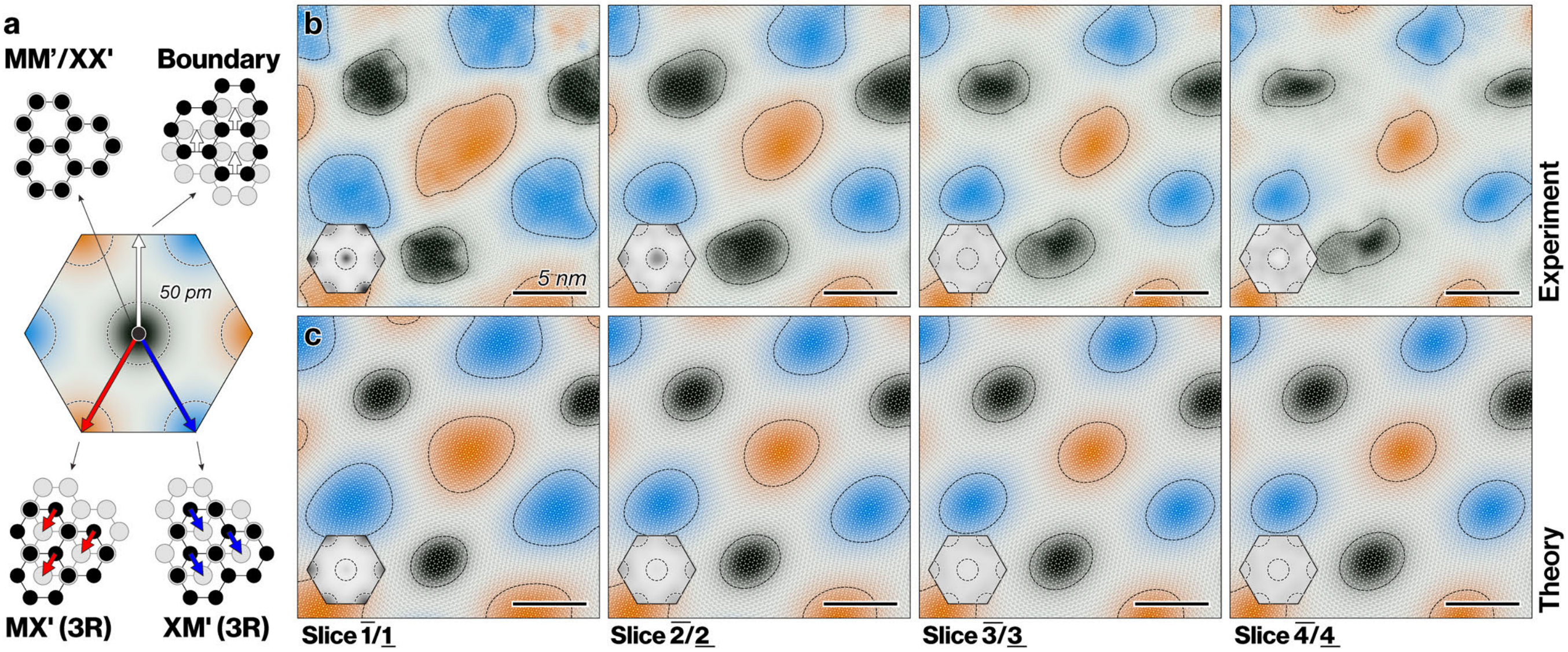

**Figure 2: Quantifying the relative displacement vector as a function of distance from the twisted interface (three $MoS_2$ layers per slice, $\theta_{twist}$ = 1.23°).** (**a**) Schematic illustrating our colour map characterizing displacement vector between $MoS_2$ layers. Zero displacement (or displacement of one unit cell) corresponds to cases where all the atomic columns and hexagonal centres are superimposed. Full displacement (solid red and blue colours) corresponds to displacement equal to the nearest neighbour distance in the high symmetry directions. (**b**) Local displacement vector extracted from pairs of reconstructed phase slices equidistant from the twisted interface, coloured according to the map in (a). The growth of the blue and red areas and compression of the black area allows us to identify the nature of the interface as parallel 3R stacking. (**c**) Local displacement vectors extracted from simulated image slices from an atomic model generated according to our continuum model for a parallel stacked interface. The insets in (b-c) show the relative distribution of different stacking areas across the field of view. Equivalent plots for an antiparallel interface are shown in Fig. S9 in the supplementary information. Relative displacement vector contours corresponding to 50 pm offsets from a high-symmetry stacking are indicated on all panels.

## Depth-dependence studies

For theoretical analysis of the depth dependence within the reconstruction, we implemented a multiscale model that accounted for stacking-dependent adhesion energy between neighbouring layers[50-52] and elastic energy created by in-plane lattice deformations.[2,53,54] This model consisted of the optimization of the total energy of the system in a moiré supercell by a continuous 3D relaxation of individual layers (see Methods and Supplementary Section S2 for details). To reproduce the experiment, we considered two bulk crystals of 2H-$MoS_2$ stacked with a twist angle of ~1.2° that formed a rhombohedral interface. The calculated relative displacement between equivalent layers on the opposite sides of the twisted interface (Fig. 2c) shows good agreement with the experimental results. Both experimental and modelling data showed that the ($\bar{4}$/$\underline{4}$) slice pair (~7 nm from the interface) was a fully rigid moiré pattern with no lattice reconstruction.

We performed a detailed analysis of the reconstruction decay within the specimen depth by using a "two-layer" slicing approach for both an $MoS_2$ homointerface (Fig. 3) and an $MoS_2/WS_2$ heterointerface (Fig. S13, S28), with the 0th slice containing one layer on either side of the twisted interface (shown schematically in Fig. 1b). Individual slices are presented in Supplementary Videos 1-6. An overlay of $\overline{1}$ and $\underline{1}$ slices ($MoS_2$ layers 2, 3, -2 and -3), is shown in Fig. 3a. These data showed atomic reconstruction to form perfectly stacked domains with all characteristic stacking regions present. The outermost slices $\overline{5}$ and $\underline{5}$ exhibited no noticeable reconstruction and could be fit with a "rigid" hexagonal lattice pattern to establish a "non-reconstructed" reference with no local image noise, to which other atomic positions could be compared. We could then calculate the displacement vector map for each individual metal atom, plotted for the $\overline{1}$ and $\underline{1}$ slices in Fig. 3b and for additional slices in Fig. S20 (see methods for the details of atomic position identification).

To describe the decay of the reconstruction with distance from the interface, we calculated the root-mean-squared (rms) displacement across the field of view for each depth section (Fig. 3c). The interfacial reconstruction effect was still present in the $\overline{2}$ and $\underline{2}$ slices with the lateral RMS displacements measured >6 pm. By the $\overline{4}$ and $\underline{4}$ slice, the RMS signal plateaued at ≈3 pm, representing the inherent uncertainty in the measured atomic positions. The observed trend in decay of the interfacial reconstruction generally agreed with our theoretical analysis, shown as black curve in Fig. 3c. To further quantify the in-plane reconstruction, we focused on a region indicated by the rectangular dashed outline in Fig. 3a, spanning across a partial dislocation between inverted 3R (XM' and MX') domains. The evolution of this region through the depth of the sample is shown in Fig. 3d. The corresponding displacement vectors as a function of distance from the twisted interface are plotted in Fig. 3e (using the same colour map as in Fig. 2a). The area of the commensurate perfectly MX' and XM' stacked regions gradually shrank, and the less favourable stackings expanded. A smooth variation of atomic registry, characteristic of a "rigid" moiré, occurred ~5 nm away from the twisted interface. By plotting the square of the local offset distance from perfect 3R stacking in Fig. 3f, we can quantitatively compare the dislocation geometry and width (Fig. 3g) between our multiscale model and that found experimentally, which demonstrated good agreement.

To estimate the depth resolution, we plot the reconstructed phase as a function of depth at the locations of atomic columns which are present in one crystal but well laterally separated from those in the other crystal (as illustrated in Fig. S17, S24, S32) in Fig. 3h. By fitting an error function to the averaged step function, we obtain a fit corresponding to a standard deviation of the underlying gaussian point-spread function, $\sigma_{PSF}$, as low as 0.94 ± 0.11 nm (corresponding to 1.26 nm using the FW80M metric). The lateral resolution was estimated by fitting Gaussian functions to our atomic columns, with $\sigma_{PSF}$ values of 30-40 pm (see SI). The depth resolution naturally means that each slice contains small contributions from surrounding layers. To understand the effect of this finite depth resolution on our analysis, we performed complementary simulations to visualise the effects of the depth blurring on our obtained results, as described in supplementary information section S1. We performed two sets of simulations, those assuming interfacial lattice reconstruction we predict theoretically ('Sim. Model'), and those assuming rigid crystals ('Sim. Rigid'). The former would act as a comparison to our experimentally resolved fields, while the latter acted as a control; any identified displacement fields here would be indicative of an artefact due to the limited depth resolution. We found that the artefacts introduced by limited depth of field were relatively small, and we could mitigate the effects by selecting only non-overlapping (>60 pm) atomic columns belonging to each crystal within the interface slices (effectively de-coupling the two interface layers), as described in supplementary information section S1. Equivalent extracted results from our simulations are plotted in Fig. 3c and 3g, and throughout the supplementary information.

We expanded the applicability of our results by conducting multiscale simulations that considered various twist angles. The atomic reconstruction for bulk crystals containing internal twisted interfaces can be summarised into three extreme cases. For small twist angles (<0.4°, see. Fig. S5a), the reconstruction propagated throughout the multilayer slab with the dislocation network gradually "blurring" in each consecutive layer. The outer surfaces still exhibited pronounced structural domains with their properties strongly affected by the complex strain texture. For the intermediate twist angles (~1-3°) the adhesion energy gain was still sufficient to trigger reconstruction in the inner layers next to the interface, but the effect dissipated within the first few layers and the outer surfaces were essentially non-reconstructed "rigid" lattices (Fig. S5). For large interfacial twist angles (>4°) the reconstruction did not occur at any depth.

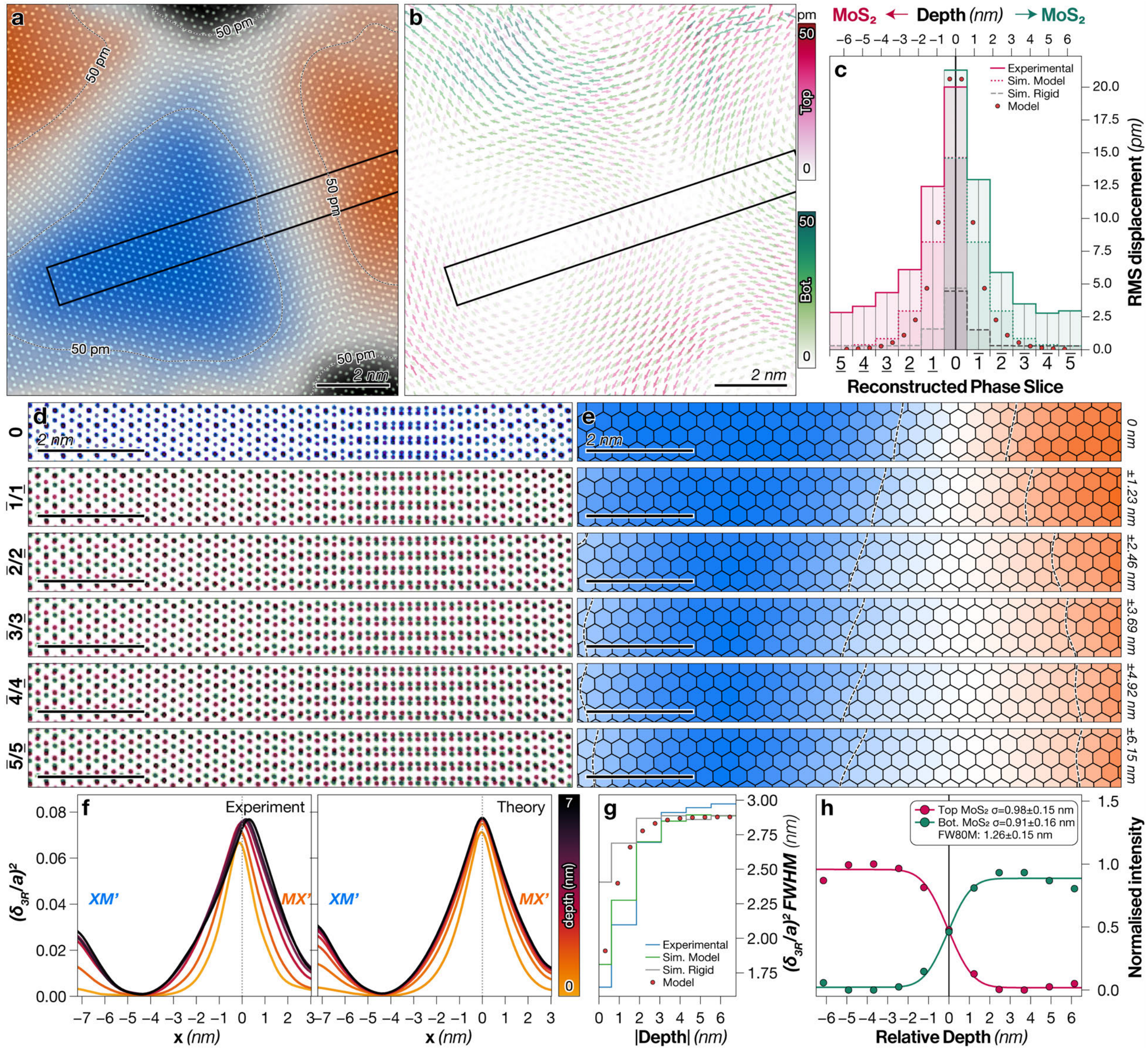

**Figure 3: Atomically resolved, reconstruction-induced displacement mapping at a $MoS_2$ homo-interface ($\theta_{twist}$ = 1.23°).** (**a**) Interface phase slice coloured by local stacking according to the scheme in Fig. 2a. (**b**) Reconstruction induced displacement of atomic columns at the interface. (**c**) RMS of measured displacement field for each two-$MoS_2$-layer slice over the field of view in (a), compared to that extracted from simulated atomic models with and without atomic reconstruction. Overlaid is the theoretically predicted RMS displacement. (**d**) Overlaid phase slices for: top row – the twisted interface, below: overlaid slices equidistant from the interface, plotted using separate colourmaps, with identified atomic positions overlaid. The area corresponds to the black box in (a). (**e**) Displacement vectors extracted from the corresponding slices in (d) coloured by local stacking according to the scheme in Fig. 2a. Each hexagon represents the offset between one hexagonal centre in one slice and the nearest hexagonal centre in the other slice. (**f**) shows the relative stacking offset from 3R stacking averaged along the long axis of the ROI in (a), compared to that predicted by our continuum model, and (**g**) shows the dislocation width obtained by measuring the FWHM of this quantity as a function of distance from the interface. (**h**) shows the intensity of isolated overlapping columns per layer providing an estimate of the effective depth resolution. In (a, e) the dashed contour lines correspond to a stacking offset of 50 pm from any high symmetry stacking.

**Strain at partial dislocations**

The reconstructed MX' and XM' rhombohedral domains identified in Figs. 1-3 are separated by partial dislocations, across which the interlayer registry transitions between equivalent inverted stackings. These boundaries are predicted to host screw-character partial dislocations in the parallel-stacked geometry, with the relative motion of the two crystals locally accommodated by shear rather than dilation.[24] By tracking the individual atomic columns to a precision of a few pm, we were able to extract the depth resolved strain tensor, ε, across an individual partial dislocation from the experimental displacement fields. Fig. 4b shows the interface-slice shear field for a 1.14° twisted $MoS_2/WS_2$ heterointerface extracted from the derivatives of the atomic displacement vectors. Equivalent plots for an additional $MoS_2/WS_2$ dataset, as well as an $MoS_2$ homostructure are included as Figs. S26 and S34 in the supplementary information. The lightness encodes the shear strain magnitude and the colour its principal direction. The MX'/XM' domains appear as approximately shear-free regions, with shear strain appearing at the 3R boundaries at the interface oriented 45° to the boundary direction. Figure 4c plots the per-crystal local shear γ, and rotation ω fields per slice, from -2.5 to 2.5 nm compared to the interface. Both quantities are progressively weaker further from the boundary, as the interfacial reconstruction decays into the bulk. The corresponding shear and rotation fields as predicted by our multiscale model (see methods and supplementary section S2) are plotted for comparison, reproducing both the spatial structure and depth decay observed. Fig 4d plots the strain components in a rotated reference frame shown in white in Fig. 4b, with x' perpendicular to a 3R boundary and quantities averaged along y' to enable a 1D profile to be presented. We compare our experimental results to both theoretical predictions, and the results extracted from our simulations (with and without atomic reconstruction (Fig. 4d). In the experimental phase data, we observe a strong peak in the $\varepsilon_{x'y'}$ profile at the boundary in opposite directions for the two crystals, and minimal variation in $\varepsilon_{x'x'}$ and $\varepsilon_{y'y'}$, the expected signature for a screw type dislocation. This behaviour is also observed in the theoretically predicted displacement data and in the simulated phase data based on theoretical values for the reconstructed atomic coordinates (termed Sim. Model, see SI for full simulation methods). The rigid crystals simulations (Sim. Rigid) when analysed with the same approach yield no displacement (as expected), giving confidence in the analysis by confirming that the experimental resolution limitations and ptychographic process do not introduce artefacts that could be mistaken for atomic reconstruction. In fact, we found the experimental data was a closer match to the purely theoretical displacement field (Theory) than when the depth resolution limitation is included (Sim. Model), demonstrating that the process reduced rather than enhanced the strength of atomic reconstruction that was measured at the interface.

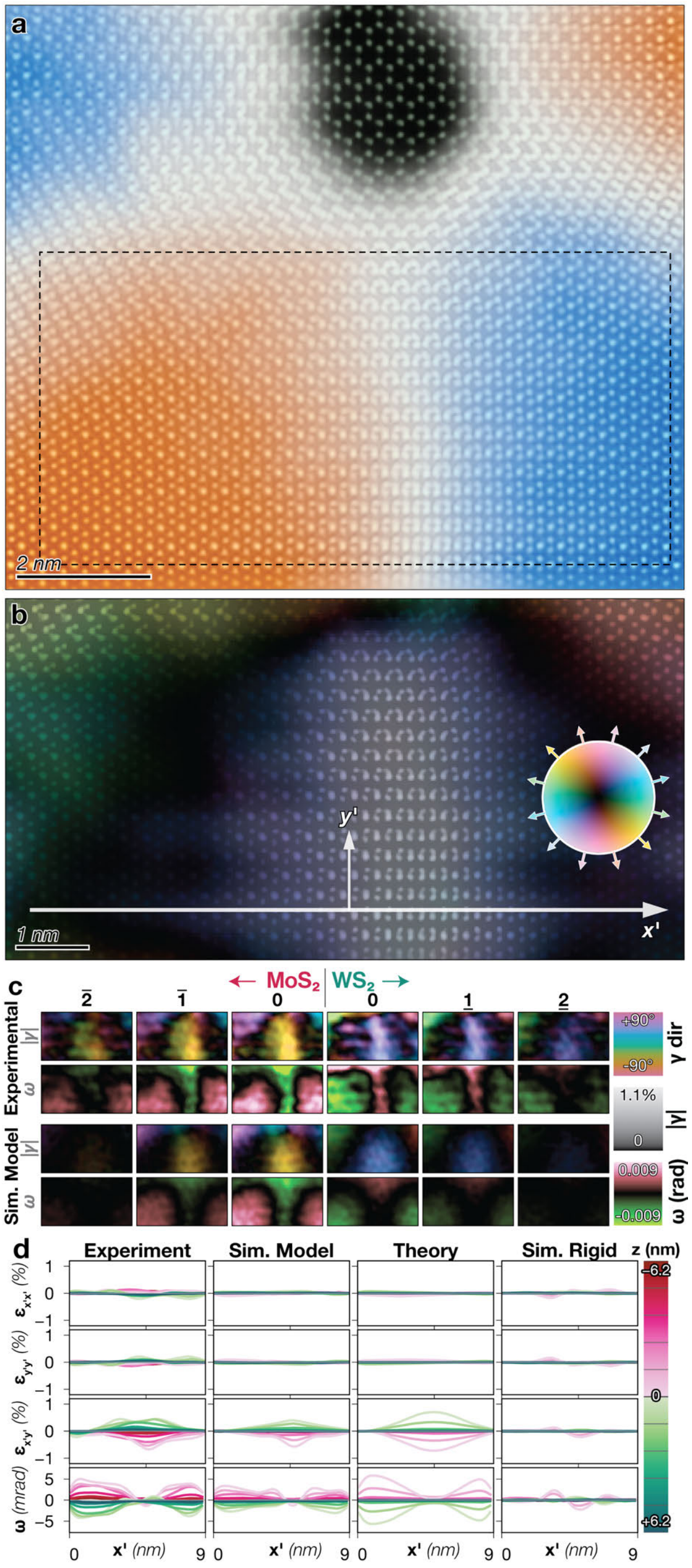

**Figure 4: Strain mapping across an MX'/XM' boundary at a 1.14° rotated $MoS_2/WS_2$ heterointerface.** (**a**) Interface phase slice coloured according to the scheme in Fig. 2a. (**b**) Interface slice shear field extracted from the atomic displacement maps (slice 0). The brightness encodes the maximum shear strain magnitude, and the hue encodes its principal direction. The area is shown by the dashed line in (a). (**c**) Maps of shear and rotation extracted from the per-crystal displacement fields from slices near the boundary compared to theoretically predicted fields. Volumetric strain was minimal as predicted theoretically. (**d**) Profiles of rotated strain tensor components averaged along a path perpendicular to a dislocation extracted from the experimental displacement maps, simulated displacement maps on reconstructed and non-reconstructed atomic models, and the underlying displacement field used to build the reconstructed atomic model, showing the screw-dislocation character. Reconstruction corresponds to 2 TMD layers per slice.

## Discussion

We show that ptychography can characterise buried interfaces of solids and can acquire detailed information on the depth dependence of atomic reconstruction of twisted interfaces for van der Waals materials. Such reconstruction appeared to be stronger than expected theoretically based on the input parameters obtained from previous bilayer studies. The 3D atomic data presented here was essential for both theoretical modelling and experimental realization of the mechanical, optical, electronic properties achievable with multilayer twisted heterostructures. For instance, correlating information about the structural properties of such interfaces with vertical (out-of-plane) transport characteristics could transform design principles for these systems. A rigidly twisted interface would present an obstacle for the propagation of Q-point electrons in the conduction band of transition metal dichalcogenides, whereas the formation of domains at the interface would enable electrons to adiabatically avoid lattice-mismatched areas occupied by the dislocations.

In mechanical and tribological studies, the emergence of the dislocation network has been attributed to abrupt switching between the structural superlubricity and high friction regime,[55] but atomic experimental data on this change has not been obtained because the critical interfaces are buried inside the structure. Correlating atomic displacements with the local frictional coefficients could enable precise atomic positioning to drive future progress in the field.

Rhombohedral interfaces are interesting for electronics because of their sliding ferroelectric properties[11] and their prospect as material systems for neuromorphic computing devices,[56] where thicker and more robust crystals can address current issues with homogeneity and reproducibility. More broadly, the power of electron ptychographic imaging to visualise and identify dislocation networks buried within relatively thick specimens can drive changes across multiple other fields such as colour centres in semiconductors, structure of heterogeneous interfaces, dislocations and many more. As computational resources become more widespread and ptychographic algorithms improve, we foresee it becoming a commonly used tool for non-invasive recovery of volumetric atomic data.

**Funding:** This work was supported by the Engineering and Physical Sciences Research Council (EPSRC) for funding under grants EP/S021531/1, EP/V001914/1, EP/P009050/1, EP/X041204/1, EP/S030719/1, EP/V007033/1 and EP/Y024303/1, and the European Research Council (ERC) under the European Union's Horizon 2020 research and innovation programme (Grant ERC-2016-STGEvoluTEM-715502). Additional TEM access was supported by the Henry Royce Institute for Advanced Materials, funded through EPSRC grants EP/R00661X/1, EP/S019367/1, EP/P025021/1 and EP/P025498/1. Financial support was provided by the University of Manchester's Dean's Doctoral Scholarship.

**Acknowledgments:** Data collection was performed using instrument E02 at the electron Physical Science Imaging Centre (ePSIC) at Diamond Light Source, project reference: MG35839.

**Competing interests:** Authors declare that they have no competing interests.

## Methods

**Sample Fabrication:** TEM samples were fabricated from mechanically exfoliated crystals, exfoliated from bulk $MoS_2$ purchased from HQ Graphene (NL). Thick crystals with natural fractures were identified, and their thickness measured using AFM. Individual parts of the fractured crystals were stacked by successively picking up with a specified sample rotation between layers, using a $SiN_x$ cantilever[45] patterned with through holes to facilitate STEM imaging,[57] which was subsequently transferred to a custom TEM compatible Si/$SiN_x$ support grid. A workflow and sample images are provided in Fig. S8.

**Acquisition:** STEM data acquisition was performed using a JEOL GrandARM300F operating at 80 kV accelerating voltage with a probe convergence of 31.7 mrad, with a probe current of 49 pA. 4D STEM CBED patterns were captured using a MerlinEM direct electron detector with a Medipix 3 sensor. 4D-STEM CBED data sets for small field of view reconstruction were captured at 10 nm defocus, 20 Mx magnification and an 8cm camera length, with an electron flux of $2.9 \times 10^7$ $e^-$ $nm^{-2}$ $s^{-1}$. Larger field of view 4D-STEM CBED data sets were captured using 20 nm defocus, 10 Mx magnification and a 12 cm camera length, and an electron flux of $7.3 \times 10^6$ $e^-$ $nm^{-2}$ $s^{-1}$.

**Ptychographic processing:** All phase images were reconstructed using the PtyREX code,[58] a parallelised and GPU accelerated implementation of mixed states ePIE[46,59] modified to include a multi-slice forward model[47] and back-propagation.[60] Additional constraints were applied to the object function to ensure unity magnitude and positive phase. For two datasets presented only in the supplementary information, a low spatial-frequency attenuation constraint was applied after every iteration to suppress low-frequency noise and improve convergence; full details are provided in Section S7 in the supplementary information. Eight incoherent probe modes were included, and scan position correction was applied after the third iteration.[61] Reconstructions converged after approximately 25 iterations, with subsequent iterations showing insignificant reduction in error. Slice thickness was chosen to minimise reconstruction error within the constraints imposed by the limits of GPU memory.

Data presented in this paper was acquired in one of two setups, optimised for large field of view or high spatial resolution. For large field of view, a camera length of 0.188 m and real space probe spacing of 0.898 Å was used. Multi-slice reconstructions were performed with 9 slices of 1.84 nm thickness. For high spatial resolution, a camera length of 0.137 m, real space probe spacing of 0.449 Å was used and 16 slices with 1.23 nm thickness in the reconstruction.

**Analysis:** The individual reconstructed phase slices were high pass filtered to remove low spatial frequencies that are not of interest for this analysis. Locations corresponding to atomic columns in slices away from the interface were identified and classified using a patch-based denoising and clustering algorithm.[62,63] Approximate locations of atomic columns corresponding to each crystal at the interface were extrapolated assuming a linear fit, then all exact sites were refined using a constrained 2D gaussian fit using the unmodified reconstructed phase slices. Full details and workflow are presented in supplementary information section S1 and Fig. S1.

**Multiscale Modelling:** Theoretical predictions of layer-resolved atomic reconstruction were derived from a multiscale continuum model based on Refs.[53,54], extended here to incorporate out-of-plane relaxation. We considered two N-layer films of 2H-$MoS_2$ (or $MoS_2/WS_2$) stacked with a relative twist angle $\theta$ at either a parallel (P) or antiparallel (AP) interface. In our model the total energy of a moiré supercell comprises three contributions: (i) the adhesion energy between neighbouring layers, modelled via an analytically interpolated functional depending on both the in-plane lateral offset and the interlayer distance, with parameters set using DFT calculations from Refs. [54,64] (Table S1); (ii) the in-plane elastic energy parameterised by the Lamé coefficient $\lambda$ and shear modulus $\mu$; and (iii) the bending energy with rigidity $\kappa$ and Poisson ratio $\sigma$ [Values from,[65,66] tabulated in Table S2]. Each layer is allowed to relax independently with both in-plane and out-of-plane displacement fields, which is essential for correctly capturing the suppression of vertical reconstruction in interior layers by the cumulative bending rigidity of the stack, and the consequent enhancement of in-plane displacements far from the twisted interface. Energy minimisation was performed using the limited-memory Broyden–Fletcher–Goldfarb–Shanno (L-BFGS) algorithm on a 2D mesh over the moiré supercell, with step size 0.5 nm for $0.2° \leq \theta \leq 0.8°$ and 0.3 nm for $1° \leq \theta \leq 4°$, to a tolerance of $10^{-11}$ eV. A full description of the model and parameter values is given in Supplementary Section S2.

**Methods References:**